\documentclass[preprint,amsmath,amssymb,aps,pra,longbibliography]{revtex4-1}

\usepackage[margin=0.75in, top=1in, bottom=0.85in]{geometry}

\usepackage[final,letterspace=-0]{microtype}
\usepackage[colorlinks=true, linkcolor=blue, urlcolor=blue, citecolor=blue, anchorcolor=blue]{hyperref}

\usepackage{mathpazo} 
\usepackage{upgreek}

\usepackage{graphicx}
\usepackage{dcolumn}
\usepackage{bm}
\usepackage{helvet}

\begin{document}

\title{Hybrid guided space-time optical modes in unpatterned films}

\author{Abbas Shiri$^{1,2}$, Murat Yessenov$^{1}$, Scott Webster$^{1}$, Kenneth L. Schepler$^{1}$, and Ayman F. Abouraddy$^{1}$}
\email{Corresponding author: raddy@creol.ucf.edu}
\affiliation{$^{1}$CREOL, The College of Optics \& Photonics, University of Central Florida, Orlando, FL 32816, USA}
\affiliation{$^{2}$Department of Electrical and Computer Engineering, University of Central Florida, Orlando, FL 32816, USA}

\begin{abstract}
Light can be confined transversely and delivered axially in a waveguide. However, waveguides are lossy static structures whose modal characteristics are fundamentally determined by the boundary conditions, and thus cannot be readily changed post-fabrication. Here we show that unpatterned planar optical films can be exploited for low-loss two-dimensional waveguiding by using `space-time' wave packets, which are the unique family of one-dimensional propagation-invariant pulsed optical beams. We observe `hybrid guided' space-time modes that are index-guided in one transverse dimension in the film and localized along the unbounded transverse dimension via the intrinsic spatio-temporal structure of the field. We demonstrate that these field configurations enable overriding the boundary conditions by varying post-fabrication the group index of the fundamental mode in a 2-$\mu$m-thick, 25-mm-long silica film, which is achieved by modifying the field's spatio-temporal structure along the unbounded dimension. Tunability of the group index over an unprecedented range from 1.26 to 1.77 around the planar-waveguide value of 1.47 is verified -- while maintaining a spectrally flat zero-dispersion profile. Our work paves the way to to the utilization of space-time wave packets in on-chip photonic platforms, and may enable new phase-matching strategies that circumvent the restrictions due to intrinsic material properties.
\end{abstract}

\maketitle

Light can be guided along optical fibers or waveguides via a refractive-index contrast \cite{SalehBook07}, a photonic bandgap \cite{Russell03Science,Joannopoulos08Book}, or other confinement mechanisms \cite{Duguay86APL,Siegman03JOSAA,Almeida04OL,Siegman07JOSAB,Wang11OL,Pryamikov11OE}. From a fundamental perspective, the salient characteristics of the guided modes, such as the mode size, group index, and group velocity dispersion are dictated by the boundary conditions \cite{SalehBook07}. In contrast to fibers and waveguides that confine light in \textit{both} transverse dimensions, a thin film (or planar waveguide) provides confinement in only \textit{one} transverse dimension, with light diffracting freely along the other unbounded dimension. Nevertheless, it is attractive to rely on unpatterned films for waveguiding because patterning of some materials can be challenging (such as organics \cite{Li2014}), and the mature technology of thin-film deposition yields ultra-smooth low-loss films \cite{Rancourt96Book} compared with the higher scattering losses associated with inscribed and etched waveguides \cite{Hall81OL,Ladouceur94IEEPO,Ladouceur97JLT,Chen12ACE,Melati14AOP}. We thus envision a novel `hybrid' guided mode in unpatterned films: the field is confined by the film along one transverse dimension and is intrinsically resistant to diffraction along the other by virtue of the field structure itself.

A potential avenue to realize this vision is to exploit diffraction-free beams \cite{Durnin87PRL,Bandres04OL,Gutierrez03AJP}. However, all such beams fundamentally require \textit{two} transverse dimensions for their realization \cite{Levy16PO}. For example, an optical field conforming to a Bessel function in only one transverse dimension diffracts -- in contradistinction to its two-dimensional counterpart. In fact, Berry proved that there are no one-dimensional beams that resist diffraction; that is, propagation-invariant light sheets do not exist -- with the exception of the Airy beam whose peak traces a parabolic trajectory (i.e., a bent light sheet) \cite{Berry79AMP,Unnikrishnan96AMP,Siviloglou07OL,Siviloglou07PRL}. However, Berry's formulation presumes a \textit{monochromatic} field. Once the monochromaticity constraint is lifted, diffraction-free \textit{pulsed} beams (or wave packets) can be constructed in the form of light sheets of arbitrary profile that travel in a straight line \cite{Dallaire09OE,Jedrkiewicz13OE,Kondakci16OE}. These wave packets are propagation-invariant by virtue of their spatio-temporal spectral structure in which each spatial frequency is precisely associated with a single temporal frequency (or wavelength), and we thus refer to them as `space-time' (ST) wave packets. Therefore, ST light sheets are the unique family of optical fields that have the potential to produce hybrid guided modes in an optical film or planar waveguide.

Early examples of ST wave packets include focus-wave modes \cite{Brittingham83JAP,Reivelt00JOSAA,Reivelt02PRE}, X-waves \cite{Lu92IEEEa,Saari97PRL,Christodoulides04OE}, among other instances \cite{Grunwald03PRA,Valtna07OC,Zamboni08PRA,Parker16OE,Wong17ACSP1,Wong17ACSP2,Porras17OL,Efremidis17OL,PorrasPRA18} -- all of which were studied with \textit{both} transverse dimensions included \cite{Reivelt03arxiv,Kiselev07OS,Turunen10PO,FigueroaBook14}. Previous theoretical studies have explored the confinement of ST wave packets in multimode two-dimensional geometries including optical fibers \cite{Vengsarkar92JOSAA} and waveguides \cite{Zamboni01PRE,Zamboni02PRE,Zamboni03PRE}. Such wave packets are superpositions of waveguide modes, with each mode associated with a prescribed wavelength to guarantee rigid wave-packet propagation. Thus, they require carefully sculpted \textit{discrete} spectra, which have not been realized to date.

We recently introduced a phase-only spatio-temporal spectral modulation scheme that produces ST wave packets in the form of a light sheet in which the field is localized along one transverse dimension only \cite{Kondakci17NP,Yessenov19OPN}. The unprecedented control over the spatio-temporal structure achievable via this strategy has enabled the observation of self-healing \cite{Kondakci18OL}, arbitrary control over the group velocity \cite{Kondakci19NC,Bhaduri19Optica}, long-distance propagation \cite{Bhaduri18OE,Bhaduri19OL}, Airy wave packets that travel in a straight line \cite{Kondakci18PRL}, and even an extension to incoherent fields \cite{Yessenov19Optica,Yessenov19OL} -- thereby raising the prospect of guidance in a planar unpatterned structure.

Here we propose and realize a new class of \textit{hybrid guided ST modes} in planar optical films that propagate self-similarly under the influence of two distinct mechanisms: traditional waveguiding in one dimension by the film, and propagation-invariant ST confinement in the other extended dimension. As such, this is the first experimental observation of a \textit{guided} diffraction-free wave packet. The versatility of our approach is brought out by confining the field in films of thicknesses extending from 100~$\mu$m down to 2~$\mu$m while maintaining transverse confinement throughout. These novel hybrid guided ST modes have continuous spectra (in contrast to Refs.~\cite{Zamboni01PRE,Zamboni02PRE,Zamboni03PRE}) and are characterized by a number of unique and salutary features. First, this configuration results in low optical losses compared to those of traditional waveguides by avoiding scattering from rough surfaces in patterned structures. Second, the two confinement mechanisms are uncoupled and can be manipulated independently of each other. Third, tuning the spatio-temporal structure of the wave packet along the unconstrained dimension enables modifying the mode size and the group index (and, in principle, the group velocity dispersion) \textit{independently} of the film thickness or refractive index. Rather than an extended planar-waveguide mode of group index $\approx\!1.47$ in a 2-$\mu$m-thick, 25-mm-long silica film, we launch into this film hybrid guided ST modes that are confined in both transverse dimensions whose group index can be tuned from $\approx\!1.26$ to $\approx\!1.77$. This unprecedented tunability range in a non-dispersive material or non-resonant structure \cite{Boyd09Science,Tsakmakidis17Science} can in principle be realized over an extended bandwidth, and demonstrates that one may override the boundary conditions in the film and modify the characteristics of the guided field post-fabrication by introducing the appropriate spatio-temporal structure into the field. Finally, hybrid guided ST modes help overcome a perennial drawback of the weak localization of free-space ST wave packets by imposing tight confinement in one transverse dimension -- \textit{without} affecting the localization in the other dimension or compromising its unique propagation characteristics. These results can help establish new prospects for phase-matching of optical pulses in nonlinear media that circumvent the restrictions imposed by the intrinsic material properties, and also offers opportunities for exploiting the useful characteristics of ST wave packets in on-chip photonic platforms. 

\section*{Results}

\subsection*{Concept of a hybrid guided ST mode}

\begin{figure*}[t!]
\centering
\includegraphics[width=14cm]{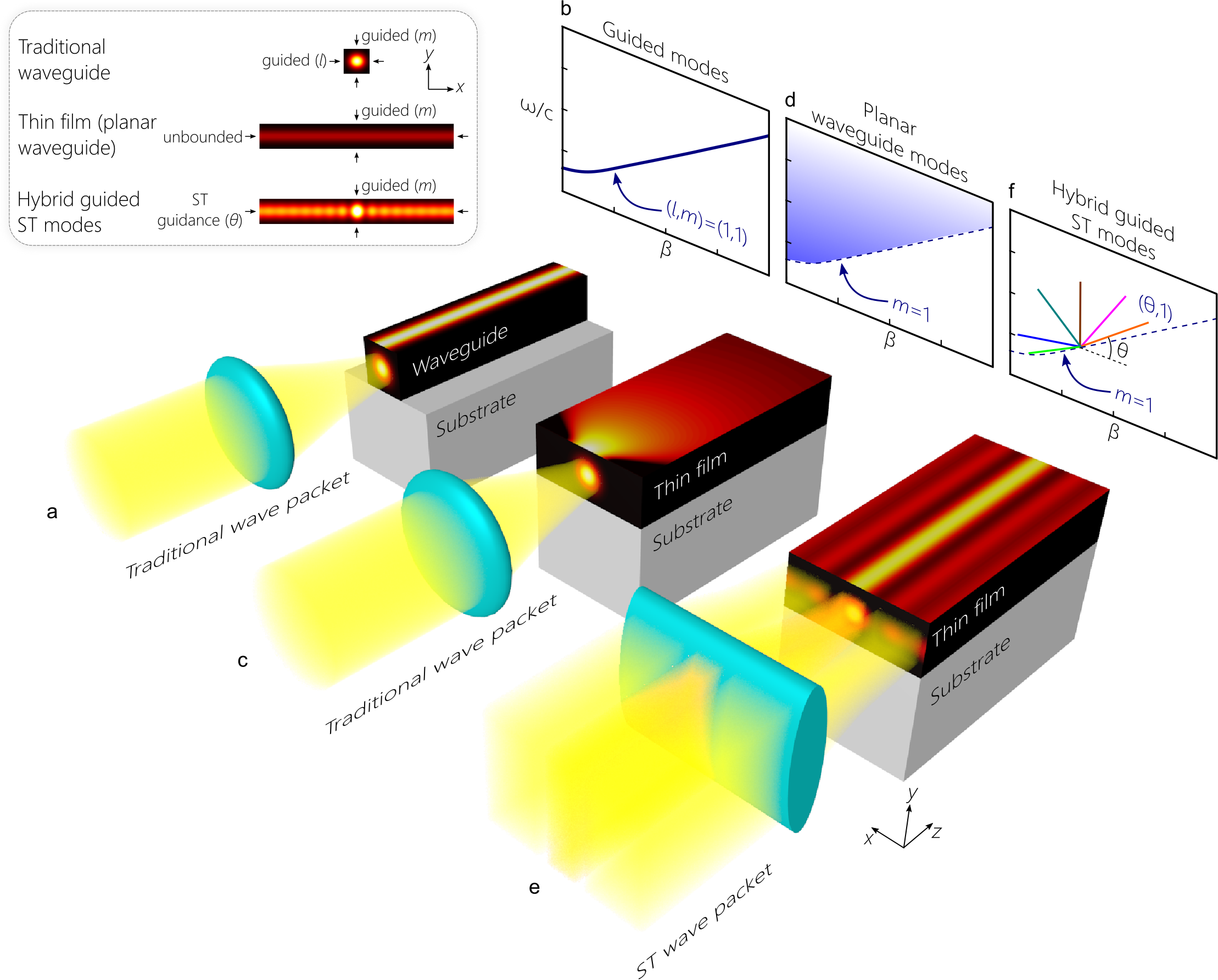}
\caption{Concept of hybrid guided ST modes. (a) Guided modes in a waveguide. A traditional optical beam is coupled into the waveguide via a spherical lens to maximize the overlap with a confined mode. (b) The dispersion relationship for a guided mode takes the form of a one-to-one mapping between the axial wave number $\beta$ and the frequency $\omega$. (c) A traditional optical beam is coupled via a spherical lens into a film. The field is confined along $y$, but diffracts along the unbounded direction $x$. (d) The dispersion relationship for each mode corresponds to an extended region in the $(\beta,\tfrac{\omega}{c})$-plane due to unbounded propagation along $x$. The field does not propagate self-similarly along the film. The dashed curve represents the limit $k_{x}\!=\!0$ for a planar waveguide extended mode. (e) A ST wave packet confined along $x$ and extended along $y$ is coupled via a cylindrical lens into a film. In addition to the confinement along $y$ by the film, the field is confined along $x$ via its intrinsic spatio-temporal structure. (f) The dispersion relationship for hybrid guided ST modes regains the one-to-one form as in (b), but its specific functional form can be tailored independently of the boundary conditions. The inset (top left corner) shows the intensity distribution of the modes and highlights the confinement mechanisms along transverse dimensions $x$ and $y$ in the three waveguiding configurations. The finite-width excitation in the planar waveguide in (c) is \textit{not} a mode.}
\label{Fig:Concept}
\end{figure*}

Traditional waveguides confine light in both transverse dimensions $x$ and $y$, as depicted in Fig.~\ref{Fig:Concept}(a). At any angular frequency $\omega$, a guided mode is associated with a \textit{single} axial wave number $\beta$. For simplicity, consider a waveguide with cross section $d\!\times\!d$, refractive index $n$, bounded by perfect mirrors, which leads to a modal dispersion relationship that takes the form $n^{2}(\tfrac{\omega}{c})^{2}-\beta^{2}\!=\!(\ell^{2}+m^{2})(\tfrac{\pi}{d})^{2}$; where $(\ell,m)$ are the discrete integer-valued modal indices along $(x,y)$ as illustrated in Fig.~\ref{Fig:Concept}(b) \cite{SalehBook07}. In effect, the boundary conditions eliminate the two transverse degrees of freedom from the dispersion relationship, and the one-to-one relationship between $\omega$ and $\beta$ thus implies the expected self-similar axial propagation of a guided mode. More generally, the boundary conditions associated with any confinement mechanism dictate a continuous differentiable dispersion relationship $\omega\!=\!f_{\ell,m}(\beta)$, that determines the modal characteristics. As one consequence of this, although a zero-dispersion condition can be realized for a guided mode at a prescribed wavelength (typically by careful design of the waveguide dimensions \cite{Turner06OE,Foster08OE}), such a condition \textit{cannot} be extended continuously over a broad bandwidth. Crucially, the modal characteristics cannot be modified post-fabrication except through thermal tuning \cite{Densmore09OE,Dong10OE} current injection in semiconductors \cite{Khan11OE}, or via a nonlinear optical effect mediated by another optical field \cite{Song05OE}. We show below that these constraints are lifted when utilizing hybrid guided ST modes.

A field focused instead into a thin film that provides confinement along only $y$ will diffract freely along the unrestricted dimension $x$ [Fig.~\ref{Fig:Concept}(c)]. For film thickness $d$, refractive index $n$, and perfectly reflecting surfaces, the dispersion relationship for the discrete set of modes indexed by a single integer $m$ is $n^{2}(\tfrac{\omega}{c})^{2}-\beta^{2}\!=\!m^{2}(\tfrac{\pi}{d})^{2}+k_{x}^{2}$, where $k_{x}$ is the component of the wave vector along $x$ (referred to subsequently as the spatial frequency). Here the boundary conditions eliminate one transverse degree of freedom from the dispersion relationship. The projected spectral representation onto the $(\beta,\tfrac{\omega}{c})$-plane is an extended domain bounded by the limit $k_{x}\!=\!0$, which corresponds to the planar waveguide extended modes [Fig.~\ref{Fig:Concept}(d)]. A one-to-one relationship between $\beta$ and $\omega$ is no longer enforced -- in contrast to the guided modes depicted in Fig.~\ref{Fig:Concept}(b), indicating that propagation along the film is \textit{not} self-similar for any localized input excitation.

Hybrid guided ST modes rely on first synthesizing a propagation-invariant wave packet in the form of a pulsed light-sheet that is extended along $y$ but \textit{non}-diffracting along $x$ \cite{Yessenov19PRA}. Focusing such a field along its extended dimension via a cylindrical lens allows matching it to a $y$-guided mode in the film while retaining the non-diffracting behavior along $x$ [Fig.~\ref{Fig:Concept}(e)]. This configuration marks a return to a one-to-one dispersion relationship in the $(\beta,\tfrac{\omega}{c})$-plane for each mode [Fig.~\ref{Fig:Concept}(f)]. The boundary conditions of the film eliminate one transverse degree of freedom, and the intrinsic spatio-temporal structure of the ST wave packets embeds the requisite additional constraint. Such a hybrid mode is indexed by $(\theta,m)$: a \textit{continuous} real parameter $\theta$ referred to as the spectral tilt angle \cite{Yessenov19OPN,Yessenov19PRA} characterizing the intrinsic field structure along $x$ and an integer $m$ for the $y$-guided modal index. Novel and useful features emerge immediately for hybrid guided ST modes, most conspicuously is that the dispersion relationship can be tailored independently of the boundary conditions of the film. For example, a purely \textit{linear} dispersion relationship can be realized (zero group velocity dispersion), and a desired group index can be realized by tuning $\theta$ without changing the modal order $m$. Uniquely, these characteristics can be realized over the full operating bandwidth of the system rather than at discrete wavelengths by sculpting the spatio-temporal spectrum of the input field, as we proceed to show.

\subsection*{ST wave packets in free space}

\begin{figure*}[t!]
\centering
\includegraphics[width=17.6cm]{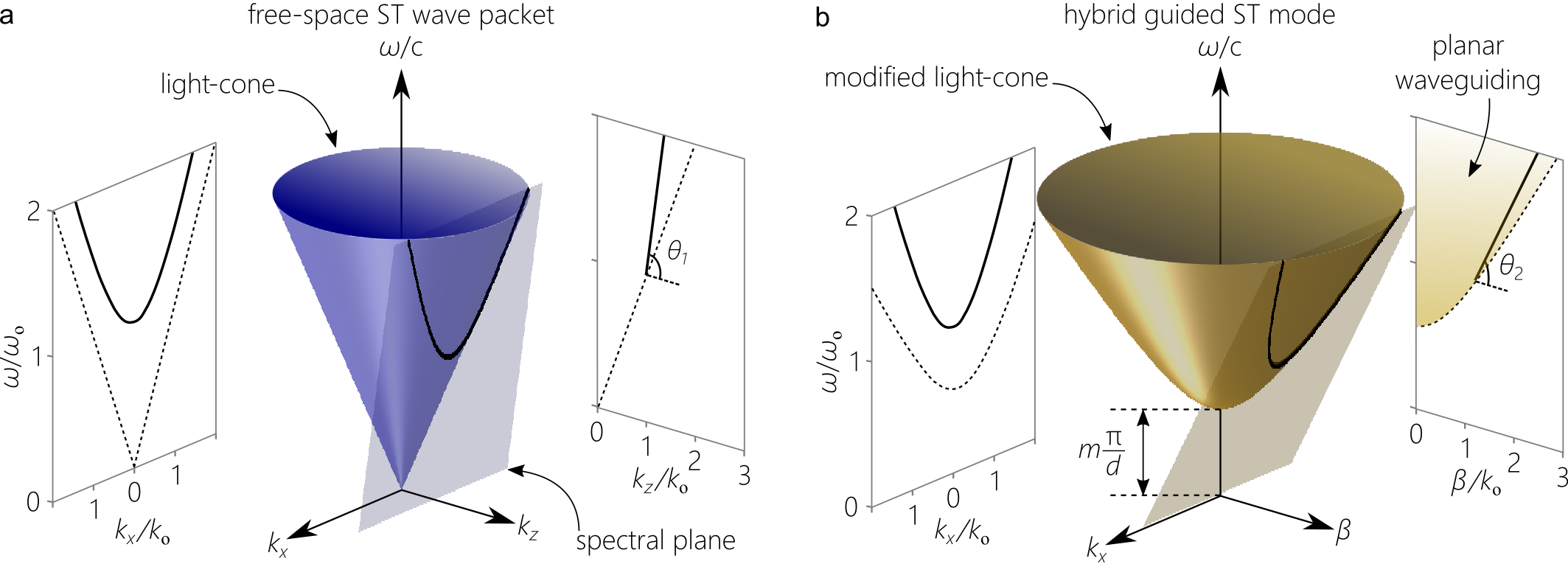}
\caption{(a) Representation of the spatio-temporal spectrum of a ST wave packet on the free-space light-cone. The spectrum lies along the intersection of the light-cone with a spectral plane tilted by an angle $\theta_{1}$ with respect to the $k_{z}$-axis. (b) Representation of the spatio-temporal spectrum of a hybrid guided ST mode in a planar film bounded by perfectly reflecting surfaces. The spectrum lies along the intersection of the light-cone associated with the $m^{\mathrm{th}}$-order mode (in a film of refractive index $n$ and thickness $d$) and a spectral plane tilted by an angle $\theta_{2}$ with respect to the $\beta$-axis. If the wave packet in (a) is to be coupled into the hybrid guided mode, the projection onto the $(k_{x},\tfrac{\omega}{c})$-plane is unchanged, and $\theta_{1}$ and $\theta_{2}$ are related through Eq.~\ref{Eq:theta}.}
\label{Fig:Cones}
\end{figure*}

To elucidate the underlying structure of the proposed hybrid guided ST modes, we first consider the propagation of their free-space counterparts. In unbounded space, the dispersion relationship $k_{x}^{2}+k_{y}^{2}+k_{z}^{2}\!=\!(\tfrac{\omega}{c})^{2}$ is satisfied by any monochromatic plane wave; here $k_{x}$, $k_{y}$, and $k_{z}$ are the components of the wave vector along the Cartesian coordinates $x$, $y$, and $z$, respectively. If we take the field to be uniform along $y$ ($k_{y}\!=\!0$), the dispersion relationship is restricted to $k_{x}^{2}+k_{z}^{2}\!=\!(\tfrac{\omega}{c})^{2}$, which corresponds geometrically to the surface of a `light-cone' [Fig.~\ref{Fig:Cones}(a)]. Creating a propagation-invariant ST wave packet in the form of a light sheet requires confining its spatio-temporal spectrum to the conic section resulting from the intersection of the light-cone with a tilted spectral plane \cite{Kondakci17NP,Yessenov19OPN} described by the equation $\tfrac{\omega}{c}\!=\!k_{\mathrm{o}}+(k_{z}-k_{\mathrm{o}})\tan{\theta}_{1}$ [Fig.~\ref{Fig:Cones}(a)]; where $k_{\mathrm{o}}$ is a fixed wave number corresponding to $\omega_{\mathrm{o}}$. This plane is parallel to the $k_{x}$-axis and makes an angle $\theta_{1}$ (the spectral tilt angle) with respect to the $k_{z}$-axis \cite{Yessenov19PRA}. Hereon, we use the subscripts `1' and `2' to indicate quantities (such as the spectral tilt angle, group velocity, and group index) for ST wave packets in free space and hybrid guided ST modes, respectively. We also make use of $k_{z}$ and $\beta$ to denote the axial wave number in free space and in the planar film, respectively. 

Uniquely, the spectral projection onto the $(k_{z},\tfrac{\omega}{c})$-plane is a \textit{straight line}, such that the group velocity along the $z$-axis is $\widetilde{v}_{1}\!=\!\tfrac{\partial\omega}{\partial k_{z}}\!=\!c\tan{\theta}_{1}$ and the group index is $\widetilde{n}_{1}\!=\!\cot{\theta_{1}}$ \cite{Kondakci19NC}. The group index is thus determined solely by $\theta_{1}$, which can be readily tuned when synthesizing the ST wave packet \cite{Kondakci19NC,Yessenov19OPN}. When $0^{\circ}\!<\!\theta_{1}\!<\!45^{\circ}$ the wave packet is subluminal ($\widetilde{v}_{1}\!<\!c$ and $\widetilde{n}_{1}\!>\!1$) and the spectral locus is an ellipse; and when $\theta_{1}\!>\!45^{\circ}$ the wave packet is superluminal ($\widetilde{v}_{1}\!>\!c$ and $\widetilde{n}_{1}\!<\!1$) and the spectral locus is a hyperbola (a parabola at $\theta_{1}\!=\!135^{\circ}$) \cite{Yessenov19PRA}. Furthermore, control over the group velocity $\widetilde{v}_{1}$ can be exercised, in principle, independently of the refractive index of the medium \cite{Bhaduri19Optica,Bhaduri19unpublished}.

\subsection*{Theory of hybrid guided ST modes}

For simplicity, we consider the above-described planar waveguide modes whose dispersion relationship for the $m^{\mathrm{th}}$-mode in a film with perfectly reflecting surfaces is $k_{y,m}^{2}+k_{x}^{2}\!=\!n^{2}(\tfrac{\omega}{c})^{2}-\beta^{2}$, which corresponds geometrically to the surface of a modified light-cone [Fig.~\ref{Fig:Cones}(b)]. Index-guiding or other confinement mechanisms result in a differently shaped light-cone, which reflects the impact of different boundary conditions. Rather than a one-to-one relationship between $\beta$ and $\omega$ for fields whose distribution along $y$ conforms to a planar-waveguide extended mode, as in Fig.~\ref{Fig:Concept}(b), the projection here takes the form of an extended domain [Fig.~\ref{Fig:Concept}(d) and Fig.~\ref{Fig:Cones}(b)]. Consequently, these localized input excitation do not propagate self-similarly.

Hybrid guided ST modes regain the one-to-one correspondence between $\beta$ and $\omega$ characteristic of traditional waveguiding by incorporating an additional constraint enforcing a one-to-one relationship between the spatial frequencies $k_{x}$ and the temporal frequencies $\omega$; that is $k_{x}\!=\!k_{x}(\omega;\theta_{2})$, where $\theta_{2}$ is a continuous real parameter. Our spatio-temporal synthesis strategy \cite{Kondakci17NP,Yessenov19OPN} allows for arbitrary relationships $k_{x}(\omega;\theta_{2})$ to be readily encoded with high precision into the field's spatio-temporal spectrum. Establishing a hybrid guided ST mode therefore requires designing the functional form of $k_{x}(\omega;\theta_{2})$ that upon substitution into the dispersion relationship of the mode reduces it to a linear form,
\begin{equation}\label{Eq:SpectralPlane}
\tfrac{\omega}{c}=k_{\mathrm{o}}+(\beta-nk_{\mathrm{o}})\tan{\theta_{2}}.
\end{equation}
This strategy is equivalent to restricting the spatio-temporal spectral support of the field to the intersection of the \textit{modified} light-cone with a tilted spectral plane defined by Eq.~\ref{Eq:SpectralPlane}, where $k_{\mathrm{o}}$ is the wave number corresponding to $\omega_{\mathrm{o}}$ on the modified light-cone, which differs slightly with respect to its free-space value. The group index $\widetilde{n}_{2}\!=\!\cot{\theta_{2}}$ of the wave packet can be tuned by tilting this spectral plane, independently of the film thickness or refractive index. Because the projection onto the $(\beta,\tfrac{\omega}{c})$-plane is now a line rather than a curved trajectory, group velocity dispersion is eliminated over the full bandwidth. 

Our approach requires first synthesizing a ST wave packet in free space in the form of a light sheet and then coupling it into the planar film. Refraction from free space to the film results in a change in the spectral tilt angle from $\theta_{1}$ to $\theta_{2}$ \cite{Bhaduri19unpublished}. Conservation of transverse momentum along $x$ and conservation of energy lead to the invariance of $k_{x}$ and $\omega$ across the planar interface of the film, and thus also the invariance of the spectral projection onto the $(k_{x},\tfrac{\omega}{c})$-plane. This results in a particularly simple expression for the refraction of ST wave packets between two materials of refractive indices $n_{1}$ and $n_{2}$ in terms of the group indices $\widetilde{n}_{1}\!=\!c/\widetilde{v}_{1}\!=\!\cot{\theta_{1}}$ and $\widetilde{n}_{2}\!=\!c/\widetilde{v}_{2}\!=\!\cot{\theta_{2}}$, whereupon $n_{1}(n_{1}-\widetilde{n}_{1})\!=\!n_{2}(n_{2}-\widetilde{n}_{2})$ \cite{Bhaduri19unpublished}. The transition from a light sheet in free space to a $y$-confined mode requires a modification of this relationship to accommodate the non-zero value of $k_{y,m}$ in the film. The new law of refraction takes the form
\begin{equation}\label{Eq:theta}
n_{1}(\widetilde{n}_{1}-n_{1})\!=\!n_{2}\left((1-\eta)\widetilde{n}_{2}-n_{2}\right),
\end{equation}
where $\eta\!=\!\tfrac{1}{2}(\tfrac{k_{y,m}}{n_{2}k_{\mathrm{o}}})^{2}$ (Methods). In our work, $n_{1}\!=\!1$ because the ST wave packet is synthesized in free space, and $n_{2}\!=\!n$ is the refractive index of the film. This relationship allows us to select the spectral tilt angle $\theta_{1}$ for the ST light sheet in free space that produces a hybrid guided ST mode with the desired $\theta_{2}$ independently of the film parameters.

Finally, we have assumed so far the idealized form of each spatial frequency $k_{x}$ related to exactly a single frequency $\omega$. Such a delta-function correlation results in infinite-energy wave packets that propagate indefinitely without change \cite{Sezginer85JAP}. In any finite system, this delta-function correlation is relaxed to a narrow but finite-width function, and thus finite-energy wave packets that propagate for large, albeit finite distances. This inevitable `fuzziness' in the association between spatial and temporal frequencies is referred to as the spectral uncertainty $\delta\omega$, which is typically much smaller than the full bandwidth $\Delta\omega$; that is, $\delta\omega\!\ll\Delta\omega$. This spectral uncertainty determines the ST wave packet propagation distance $L\!\sim\!\tfrac{c}{\delta\omega}\tfrac{1}{|1-\cot{\theta}|}$ \cite{Yessenov19OE,Kondakci19OL}.

\subsection*{Synthesis and characterization of ST wave packets}

\begin{figure}[t!]
\centering
\includegraphics[width=8.0cm]{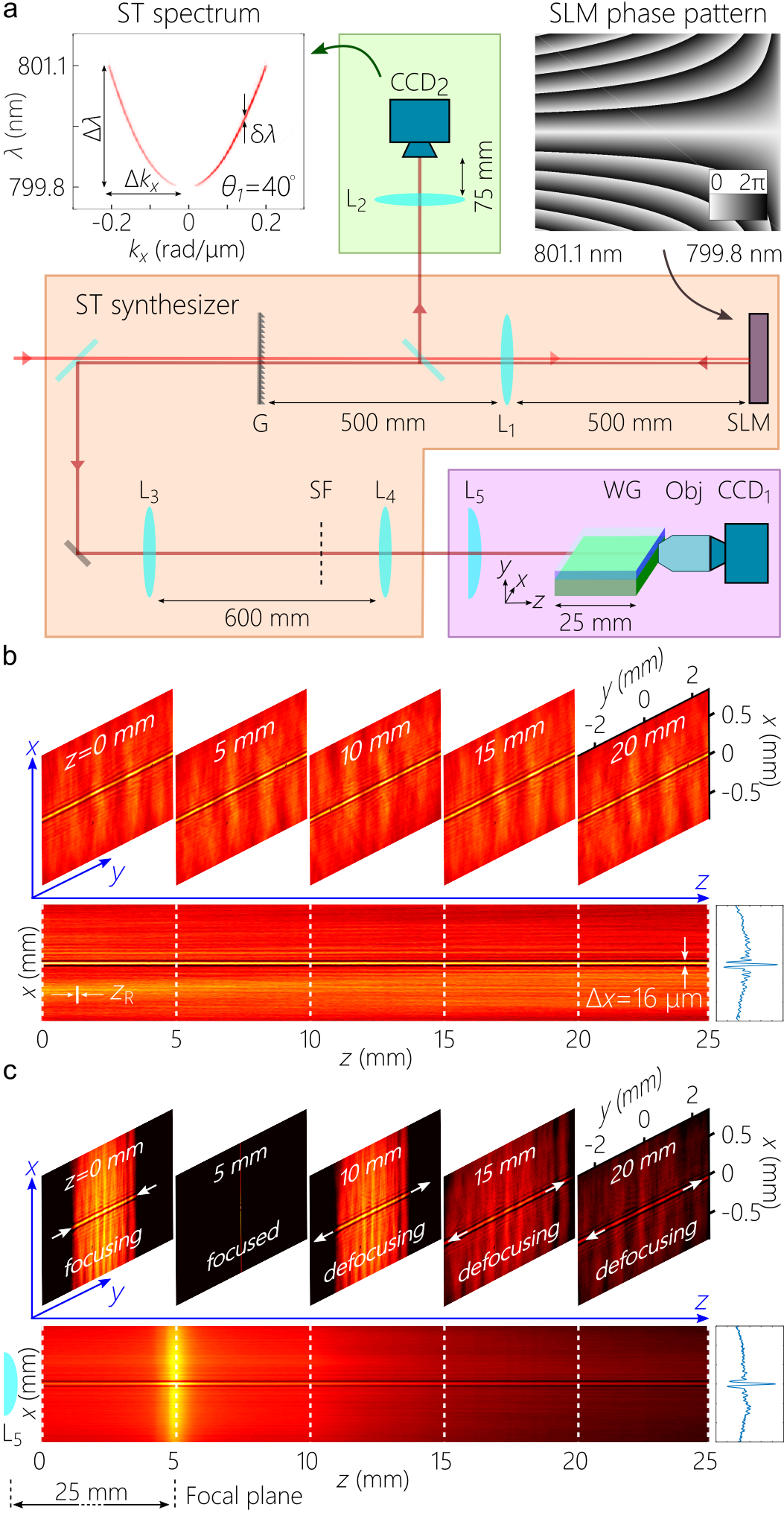}
\caption{(a) Setup for synthesis and characterization of ST wave packets, and launching hybrid guided ST modes into a film. G: Diffraction grating; L: lens; SF: spatial filter; SLM: spatial light modulator; Obj: objective lens; WG: unpatterned planar waveguide or film. Right inset is the phase imparted by the SLM for $\theta_{1}\!=\!40^{\circ}$, and left inset is the corresponding measured spatio-temporal spectrum captured by CCD$_2$. (b) Axial evolution of the time-averaged intensity $I(x,z)$ of the propagation-invariant ST wave packet (the lens L$_5$, the planar waveguide, and objective lens are removed). The transverse intensity distributions in the $(x,y)$-plane are shown at selected axial positions, emphasizing the propagation invariance. The white line is the Rayleigh range $z_{\mathrm{R}}\!\approx\!0.25$~mm for a traditional beam having a spatial width $\Delta x\!=\!16$~$\mu$m. (c) Axial evolution of the time-averaged intensity $I(x,z)$ of ST wave packet after placing a cylindrical lens (L$_5$, focal length $25$~mm) that focuses the field along $y$. The transverse intensity distributions in the $(x,y)$-plane are shown at selected axial positions; the field first contracts along $y$ at the focal plane and then diverges. The plots on the right in (b) and (c) correspond to $I(x,0)$.}
\label{Fig:Setup}
\end{figure}

The ST wave packets are synthesized using femtosecond laser pulses from a Ti:sapphire laser directed into the folded pulse-shaping setup depicted in Fig.~\ref{Fig:Setup}(a) \cite{Yessenov19OPN} (Methods). A diffraction grating disperses the pulse spectrum spatially and a cylindrical lens maps the wavelengths to positions along a spatial light modulator (SLM) that imparts to each wavelength a linear phase distribution along the direction orthogonal to that of the spread spectrum to assign a particular $k_{x}$ to each wavelength. The phase distribution imparted to the spectrally resolved wave front [Fig.~\ref{Fig:Setup}(a), right inset] introduces into the field the requisite spatio-temporal spectral structure to realize a spectral tilt angle $\theta_{1}$. The retro-reflected wave front from the SLM is reconstituted into a pulsed beam at the grating. The temporal bandwidth $\Delta\lambda$ throughout is $\approx\!1.3$~nm ($799.8-801.1$~nm). The spatial width of the ST wave packet is $\Delta x\!=\!\pi/\Delta k_{x}$, where $\Delta k_{x}$ is the spatial bandwidth. The spectral uncertainty $\delta\lambda$ (in units of wavelengths) is determined predominantly by the spatial width of the diffraction grating G, which determines the achievable spectral resolution. In our experiment, we expect $\delta\lambda\!\sim\!20$~pm. We select a range of values of the spectral tilt angle $\theta_{1}$ that guarantees a minimum propagation distance of 25~mm, which is the length of the waveguide \cite{Yessenov19OE,Kondakci19OL}. Further propagation distances can be achieved by using larger-width gratings.

To confirm that the targeted spatio-temporal spectrum is realized, we resolve the spatial spectrum of the wave front retro-reflected from the SLM, which is captured by a CCD camera (CCD$_2$). This measurement reveals the conic section relating the spatial and temporal frequencies [Fig.~\ref{Fig:Setup}(a), left inset]. The field is characterized in physical space by scanning a CCD camera (CCD$_1$) along $z$ to obtain the time-averaged intensity $I(x,z)\!=\!\iint\!dydt\,I(x,y,z;t)$, an example of which is plotted in Fig.~\ref{Fig:Setup}(b) corresponding to $\theta_{1}\!=\!40^{\circ}$. The diffraction-free behavior along $z$ is clear.

\subsection*{Launching of hybrid guided ST modes}

The synthesized ST wave packets are in the form of $y$-polarized (TM-polarized) light sheets that are uniform along $y$ and structured along $x$. To launch these light sheets into a film, we focus the field with a cylindrical lens (L$_5$) along $y$, and the axial evolution of the intensity $I(x,z)$ shows that the propagation invariance displayed by the unfocused wave packet [Fig.~\ref{Fig:Setup}(b)] is retained after focusing [Fig.~\ref{Fig:Setup}(c)] -- except for the abrupt increase at the axial plane corresponding to the lens focus. The field converges along $y$ to reach a minimal width at this plane, and then subsequently diverges along $y$, all the while maintaining the same field distribution along $x$. That is, focusing does \textit{not} adversely impact the diffraction-free behavior along $x$. The input facet of the film is placed at the focal plane of the lens L$_5$. 

The films used in our experiments consist of a layer of SiO$_2$ ($n\!\approx\!1.46$ at $\lambda\!=\!800$~nm) of thicknesses 2~$\mu$m and 5~$\mu$m deposited via e-beam evaporation onto a MgF$_2$ substrate (MSE Supplies; $n\!\approx\!1.38$ at $\lambda\!=\!800$~nm) of area $25\times25$~mm$^{2}$. The film confines the field along $y$ via index-guiding with respect to air cover and the MgF$_2$ substrate, whereas ST guidance maintains the field distribution invariant along $x$. In addition, we make use of 100-$\mu$m-thick glass microscope slides (Thorlabs CG00C2) of area $22\times22$~mm$^{2}$. After coupling the field into the film, the output field is imaged through an objective lens to CCD$_{1}$. Once the field is coupled into the waveguide, no alignment changes are necessary when changing the spectral tilt angle $\theta_{1}$, which requires only sculpting the phase implemented by the SLM.

\subsection*{Observation of hybrid guided ST modes}

\begin{figure}[t!]
\centering
\includegraphics[width=8.8cm]{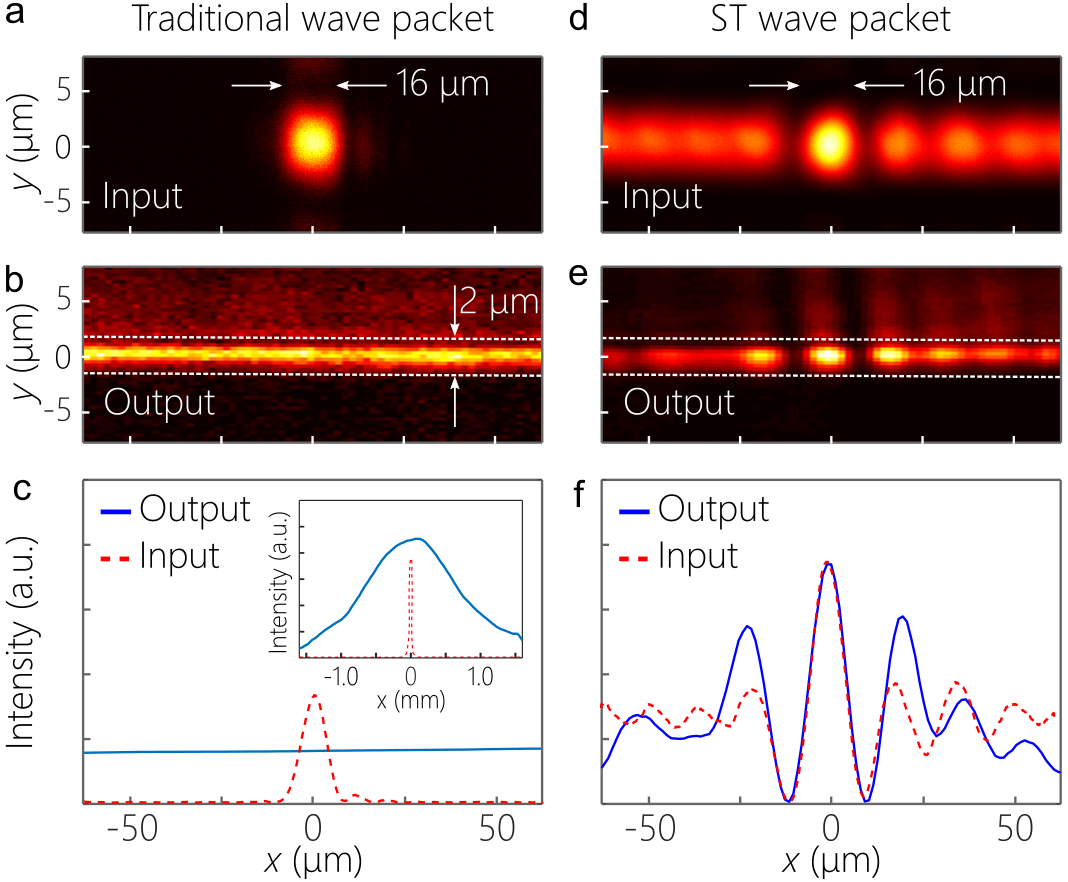}
\caption{Observation of hybrid guided ST modes. The planar waveguide consists of a 2-$\mu$m-thick layer of SiO$_2$ on a MgF$_2$ substrate. (a-c) Propagation of a traditional Gaussian wave packet with separable spatial and temporal degrees of freedom in the film. (a) Intensity at the input and (b) at the output of the film. (c) One-dimensional distribution through the center of (a) and (b). Inset shows the plot on a wider horizontal scale. (d-f) Same as (a-c) for an ST wave packet at the input ($\theta_{1}\!=\!40^{\circ}$) having the same width as the Gaussian wave packet in (a). In contrast to the traditional Gaussian wave packet, here the input and output intensity profiles are almost identical.}
\label{Fig:STandGaussian}
\end{figure}

To establish the propagation-invariance of hybrid guided ST modes in a film, we compare their propagation to a traditional Gaussian wave packet along the film whose spatial and temporal degrees of freedom are separable. We plot the intensity at the entrance and exit of the 2-$\mu$m-thick, 25-mm-long film for a traditional Gaussian wave packet [Fig.~\ref{Fig:STandGaussian}(a-c)] having $\Delta x\!\approx\!16$~$\mu$m (Rayleigh range of $\sim\!0.25$~mm), and for a ST wave packet of equal transverse width [Fig.~\ref{Fig:STandGaussian}(d-f)] corresponding to $\theta_{1}\!=\!40^{\circ}$ and $\Delta k_{x}\!\approx\!0.2$~rad/$\mu$m. The hybrid guided ST mode profile remains unchanged, whereas diffractive spreading broadens the field along $x$ for the Gaussian wave packet. Comparison between the input and output intensity distributions in these two cases is facilitated by examining one-dimensional sections along $x$ for the Gaussian wave packet [Fig.~\ref{Fig:STandGaussian}(c)] and hybrid guided ST mode [Fig.~\ref{Fig:STandGaussian}(f)]. The output intensity for the Gaussian wave packet is almost flat over the relevant transverse scale (the width at the output has extended to $\sim\!1.6$~mm; see Fig.~\ref{Fig:STandGaussian}(c) inset). These measurements establish that the hybrid guided ST mode is bound in both dimensions in the planar waveguide for propagation distances extending to at least 25~mm. 

\begin{figure}[t!]
\centering
\includegraphics[width=8.8cm]{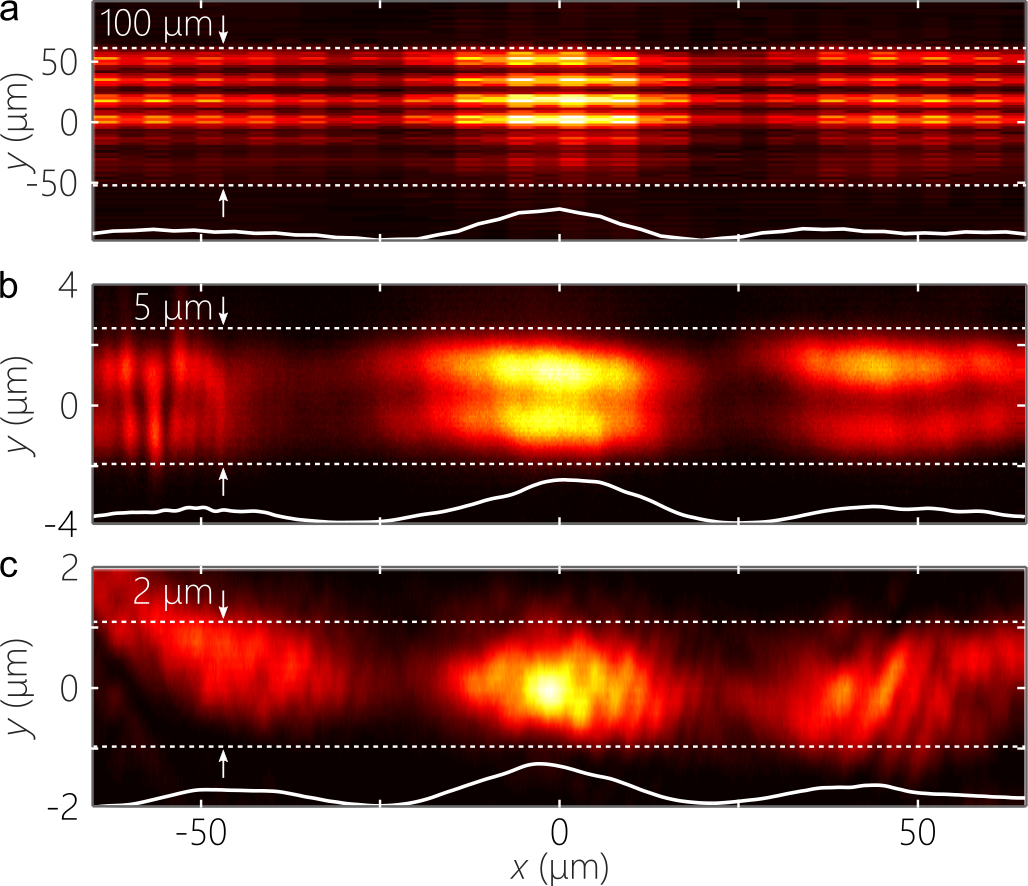}
\caption{Hybrid ST modes in films of thickness (a) 100~$\mu$m (microscope slide), (b) 5~$\mu$m (SiO$_2$ on MgF$_2$), and (c) 2~$\mu$m (SiO$_2$ on MgF$_2$). The three cases correspond to $\theta_{1}\!=\!44^{\circ}$, $\Delta x\!\approx\!25$~$\mu$m. The white curves are integrated over $y$, and are almost identical in the three cases.}
\label{Fig:WaveGuideSize}
\end{figure}

As mentioned earlier, a key feature of hybrid guided ST modes is that their characteristics can be tuned independently in the two transverse dimensions. We show a first example of this versatility by coupling ST wave packets into planar waveguides of thicknesses 100~$\mu$m [Fig.~\ref{Fig:WaveGuideSize}(a)], 5~$\mu$m [Fig.~\ref{Fig:WaveGuideSize}(b)], and 2~$\mu$m [Fig.~\ref{Fig:WaveGuideSize}(c)]. The free-space ST wave packet coupled into the films has a spectral tilt angle of $\theta_{1}\!=\!44^{\circ}$ and a spatial bandwidth $\Delta k_{x}\!=\!0.12$~rad/$\mu$m, corresponding to a transverse width of $\Delta x\!=\!25$~$\mu$m. Despite the wide variation of film thickness $d$ in these three cases, measurements at the output show that the \textit{same} width of the guided mode is nevertheless maintained along $x$, whereas the modal structure along $y$ is dictated by the film thickness $d$.

\subsection*{Tuning the group index of a hybrid guided ST mode}

We next proceed to verify a salient characteristic of the hybrid guided ST mode, namely, that the group index $\widetilde{n}_{2}$, can be tuned independently of the boundary conditions of the film by tailoring the intrinsic spatio-temporal spectral structure of the field. To demonstrate this, ST wave packets are coupled into a 2-$\mu$m-thick film while varying $\theta_{1}$ by changing the SLM phase distribution. Because the spatial and temporal bandwidths are related by a factor that depends on the spectral tilt angle \cite{Yessenov19OE}, fixing $\Delta\lambda$ here while varying $\theta_{1}$ results in a concomitant change in $\Delta k_{x}$ (and hence in transverse width $\Delta x$). Maintaining $\Delta x$ while varying $\theta_{1}$ requires changing $\Delta\lambda$. In Fig.~\ref{Fig:ChangingTheta} we plot measurements of the spatio-temporal spectral intensity and the intensity distributions at the input and output of the film for $\theta_{1}\!=\!35^{\circ}$, $40^{\circ}$, $44.5^{\circ}$, $45^{\circ}$ (plane-wave pulse), $45.5^{\circ}$, $50^{\circ}$, and $55^{\circ}$; corresponding to spectral tilt angles in the film of $\theta_{2}\!=\!29.5^{\circ}$, $32.0^{\circ}$, $34.0^{\circ}$, $34.2^{\circ}$, $34.4^{\circ}$, $36.3^{\circ}$ and $38.4^{\circ}$, respectively. In all cases, the output distribution along $x$ after the 25-mm-long film is almost identical to the input. The minimum spatial width along $x$ is $\Delta x\!\approx\!10$~$\mu$m at $\theta_{1}\!=\!35^{\circ}$. At $\theta_{1}\!=\!45^{\circ}$ ($\theta_{2}\!=\!34.2^{\circ}$ and $\Delta k_{x}\!=\!0$), light is coupled into a traditional planar waveguide extended mode whose group index $n_{\mathrm{g}}\!=\!1.47$ is determined solely by the boundary conditions in the film. The first three modes are subluminal; that is, $\widetilde{v}_{2}\!<\!c/n_{\mathrm{g}}$ and $\widetilde{n}_{2}\!>\!n_{\mathrm{g}}$, where $\widetilde{n}_{2}$ is the group index of the hybrid guided ST mode obtained via Eq.~\ref{Eq:theta}. The last three modes are superluminal with $\widetilde{v}_{2}\!>\!c/n_{\mathrm{g}}$ and $\widetilde{n}_{2}\!<\!n_{\mathrm{g}}$. The transition from subluminal to superluminal modes is associated with a switch in the sign of the curvature of the conic section representing the spatio-temporal spectral intensity. 

The tunability of the group index $\widetilde{n}_{2}$ of the hybrid guided ST mode can be appreciated when compared to the planar waveguide extended modes (associated with plane-wave input $\Delta k_{x}\!=\!0$). The dispersion relationships for the lowest-order index-guided TM modes in the film are plotted in Fig.~\ref{Fig:2um_singlemode}(a), which are determined solely by the boundary conditions. In the spectral range of interest, the fundamental TM mode has a group index of $n_{\mathrm{g}}\!=\!1.47$, whereas the index for SiO$_2$ is $n\!=\!1.46$, indicating the usual `slowing' of a pulse in an index-guided structure. In Fig.~\ref{Fig:2um_singlemode}(b) we plot the projection of the spatio-temporal spectrum for the hybrid guided ST modes onto the $(\beta,\lambda)$-plane for the 7 realizations in Fig.~\ref{Fig:ChangingTheta} (Methods). Note that the dispersion relationship is in each case a \textit{straight line} (zero group velocity dispersion across the banwidth utilized) that makes a spectral tilt angle $\theta_{2}$ with respect to the $\beta$-axis. Although we make use of the fundamental TM mode for the index-guided film over our operating spectral window in all cases, the group index can nevertheless be tuned by varying the spectral tilt angle. Measurements reveal $\widetilde{n}_{2}$ in the range from $\approx\!1.26$ (superluminal, $\widetilde{n}_{2}\!<\!1.47$, $\widetilde{v}_{2}\!>\!c/1.47$) to $\approx\!1.77$ (subluminal, $\widetilde{n}_{2}\!>\!1.47$, $\widetilde{v}_{2}\!<\!c/1.47$). In other words, hybrid guided ST modes can be made to travel along the film faster or slower than the extended modes of the planar waveguide sharing the same modal index $m$. This wide tunability range is unprecedented in a fixed structure consisting of nondispersive materials and demonstrates how the spatio-temporal structure of the hybrid guided ST modes helps override the boundary conditions. Furthermore, a wider range of potential modal group indices are -- in principle -- accessible and broader operational bandwidths can be implemented \cite{Kondakci18OE}.

\subsection*{Propagation losses of hybrid guided ST modes}

Imaging the planar waveguide top surface did not reveal any out-scattered light from the hybrid guided ST mode, except at the input facet, indicating that the main source of loss is due to the coupling efficiency into the film. Measurements revealed a transmission of $\approx\!32-34\%$ into the 2-$\mu$m-thick film for spectral tilt angles $\theta_{1}$ in the range $35-55^{\circ}$ for the free-space ST light sheets (Methods). Calculated coupling efficiencies into the planar waveguide based on the measured input field structure predict values of $38-42\%$ for the fundamental TM mode. After correcting for Fresnel reflection at the interfaces, an unaccounted loss of $5-10\%$ was left that we attribute to absorption and scattering in the film. This leads to an \textit{upper} limit on the transmission losses of $0.1-0.2$~dB/cm along the film, which is very low compared to traditionally inscribed or etched waveguides where substantially higher losses result from wall roughness.

\section*{Discussion}

By exploiting the unique family of propagation-invariant ST wave packets in the form of light sheets, we have demonstrated for the first time the utility of diffraction-free fields in guided optics and potentially in on-chip platforms. In our work, the optical field is guided in the thin film through a hybrid of two mechanisms: index-guiding in the direction normal to the film (other reflection mechanisms can in principle be utilized), and in the direction parallel to the film via the intrinsic spatio-temporal spectral structure of the field. Hybrid guided ST modes are thus solutions to the wave equation with boundary conditions enforced in only one dimension, and no boundary conditions needing to be satisfied along the other (in contrast to previous work \cite{Zamboni01PRE,Zamboni02PRE,Zamboni03PRE}). Hybrid guided ST modes can thus be viewed as `virtual' waveguides in unpatterned films. We have confirmed here several of the unique properties of hybrid guided ST modes. First, because waveguide fabrication is not required, low-loss transmission is realized with an upper limit of $0.1-0.2$~dB/cm. Second, the properties of the hybrid guided ST mode can be held fixed while varying the parameters of the film. Third, the hybrid guided ST mode can be tuned by modifying the spatio-temporal structure of the field. Using this strategy, we varied the group index in the range $1.26-1.77$ around the group index value of 1.47 for the extended mode of the planar waveguide. These unique features can be realized continuously over extended bandwidths. Other families of ST wave packets such as X-waves and focus-wave modes can potentially also be used to launch hybrid guided ST modes once they are synthesized in the form of light sheets (which was recently achieved in \cite{Yessenov19PRA}).

Future work will be directed to the measurement of the group delays incurred by hybrid guided ST modes traversing the film. Another exciting prospect is the study of the ST wave packets with negative group velocities \cite{Kondakci19NC,Yessenov19OE,Zapata06OL} propagating along a waveguide. Furthermore, we have recently synthesized multiple co-propagating ST wave packets in the same spatial channel while occupying the same or different spectral channels with independently addressable group velocities \cite{Yessenov19unpub}. Combined with the demonstration reported here, these results lay the foundation for constructing on-chip optical delay lines based on confined ST wave packets, which may lead to the realization of all-optical buffers that have remained elusive to date. Coupling such sophisticated field configurations into a thin film with tight confinement suggests striking opportunities in nonlinear optics and light-matter interactions. Unique phase-matching scenarios can be envisioned by controlling the group index and dispersion profile independently of the wavelength, the planar waveguide thickness, or refractive index. Finally, the successful confinement of \textit{one}-dimensional ST wave packets to a thin-film waveguide suggests an intriguing prospect: \textit{surface} ST modes; e.g., surface plasmon polaritons whose transverse confinement results from endowing the field with the appropriate spatio-temporal spectral correlations, leading to propagation-invariant plasmonic excitations that are localized in all dimensions. 

\bibliography{diffraction}

\vspace{2mm}
\noindent
\textbf{Acknowledgments}\\
We thank Sasan Fathpour for assistance with planar-waveguide polishing, Kyu Young Han for loan of equipment and useful suggestions, and  Ali K. Jahromi for helpful discussions. This work was supported by the U.S. Office of Naval Research (ONR) under contracts N00014-17-1-2458 and N00014-19-1-2192.

\vspace{2mm}
\noindent
\textbf{Author contributions}\\
\noindent
K.L.S. and A.F.A. developed the concept and supervised
the research. A.S. designed the experiment, carried out the measurements, and analyzed the data, with assistance from M.Y. and S.W. All authors contributed to writing the paper.

\noindent
Correspondence and requests for materials should be addressed to A.F.A.\\(email: raddy@creol.ucf.edu)

\vspace{2mm}
\noindent
\textbf{Competing interests:} The authors declare no competing interests.
\clearpage

\section*{Methods}

\subsection*{Derivation of the law of refraction of ST light sheets to hybrid guided ST modes}

In an extended medium of refractive index $n_{1}$, the dispersion relationship for a plane monochromatic wave when assuming the field uniform along $y$ ($k_{y}\!=\!0$) is
\begin{equation}
k_{x}^{2}+k_{z}^{2}\!=\!n_{1}^{2}(\tfrac{\omega}{c})^{2}.
\end{equation}
A ST wave packet requires satisfying an additional constraint $k_{z}\!=\!n_{1}k_{\mathrm{o}}+(\tfrac{\omega}{c}-k_{\mathrm{o}})\widetilde{n}_{1}$. In a planar waveguide of refractive index $n_{2}$, and for the $m^{\mathrm{th}}$-order mode having a $y$-component of the wave vector $k_{y,m}$, the dispersion relationship is
\begin{equation}
k_{x}^{2}+k_{y,m}^{2}+\beta^{2}\!=\!n_{2}^{2}(\tfrac{\omega}{c})^{2}.
\end{equation}
Note that $k_{x}$ and $\omega$ are invariant upon transmission across a planar interface between the two materials that is orthogonal to the $z$-axis, whereas $k_{z}$ is not. Subtracting the two dispersion relationships $k_{y,m}^{2}+\beta^{2}-k_{z}^{2}\!=\!(n_{2}^{2}-n_{1}^{2})(\tfrac{\omega}{c})^{2}$ and then substituting the constraint for the ST wave packet in the first medium eliminates $k_{z}$,
\begin{equation}
(\widetilde{n}_{1}^{2}-n_{1}^{2}+n_{2}^{2})(\tfrac{\omega}{c})^{2}-2\widetilde{n}_{1}(\widetilde{n}_{1}-n_{1})k_{\mathrm{o}}(\tfrac{\omega}{c})+(\widetilde{n}_{1}-n_{1})^{2}k_{\mathrm{o}}^{2}-k_{y,m}^{2}-\beta^{2}=0.
\end{equation}
The group velocity of the hybrid guided ST mode is $\widetilde{v}_{2}\!=\!\tfrac{\partial\omega}{\beta}$, and the group index is $\widetilde{n}_{2}\!=\!\tfrac{c}{\widetilde{v}_{2}}$,
\begin{equation}
\widetilde{n}_{2}\beta=(\widetilde{n}_{1}^{2}-n_{1}^{2}+n_{2}^{2})\tfrac{\omega}{c}-\widetilde{n}_{1}(\widetilde{n}_{1}-n_{1})k_{\mathrm{o}},
\end{equation}
which is to be evaluated at $\omega\!=\!\omega_{\mathrm{o}}$.

In absence of a waveguiding structure $k_{y}\!=\!0$, in which case $\beta\!=\!n_{2}k_{\mathrm{o}}$, and we retrieve the law of refraction from Ref.~\cite{Bhaduri19unpublished}, $n_{1}(\widetilde{n}_{1}-n_{1})\!=\!n_{2}(\widetilde{n}_{2}-n_{2})$. In the waveguide, $\omega\!=\!\omega_{\mathrm{o}}$ when $k_{x}\!=\!0$ and $\beta^{2}\!=\!n_{2}^{2}k_{\mathrm{o}}^{2}-k_{y,m}^{2}$. For $k_{y,m}\!\ll\!n_{2}k_{\mathrm{o}}$, $\beta\!\approx\!n_{2}k_{\mathrm{o}}(1-\eta)$, where $\eta\!=\!\tfrac{1}{2}(\tfrac{k_{y,m}}{n_{2}k_{\mathrm{o}}})^{2}$. Substituting for $\beta$, we obtain a modified law of refraction, $n_{1}(\widetilde{n}_{1}-n_{1})\!=\!n_{2}(\widetilde{n}_{2}(1-\eta)-n_{2})$, which we make use of in the main text.

\subsection*{Synthesis and characterization of ST light sheets}

Femtosecond pulses from a Ti:Sapphire laser (Coherent, MIRA 900) are directed to a diffraction grating G (Thorlabs GR25-1208, 1200 lines/mm, area $25\!\times25$~mm$^{2}$) that disperses the pulse spectrum spatially; see Fig.~\ref{Fig:Setup}(a). A cylindrical lens L$_1$ of focal length 500~mm maps the wavelengths to positions along a SLM (Hamamatsu X10468-02). The retro-reflected wave front from the SLM through L$_1$ is reconstituted into a pulsed beam at the grating G. Lenses L$_3$ and L$_4$ in an imaging configuration reduce the size of the beam by $\times5$. 

The spatio-temporal spectrum of the retro-reflected field from the SLM is captured by a CCD camera (CCD$_2$; The ImagingSource DMK 27BUP031) after a spherical lens L$_2$ of focal length 75~mm. The measured spatial intensity profile is a segment from a conic section centered at its axis of symmetry $k_{x}\!=\!0$. In the limit of small bandwidth $\Delta\lambda\!\ll\!\lambda_{\mathrm{o}}$, any such conic section can be approximated by a parabola [Fig.~\ref{Fig:Setup}(a), inset]. The field is characterized in physical space by scanning a CCD camera (CCD$_1$; The ImagingSource DMK 27BUP031) on a linear stage along the axial coordinate $z$ in steps of 100~$\mu$m for a distance of 25~mm. 

\subsection*{Coupling of ST wave packets into a planar waveguide}

Both facets of the 100-$\mu$m-thick microscope slide were polished to 0.1~$\mu$m with silicon carbide sandpapers. To couple the ST wave packet into any of the films used (of thicknesses 100, 5, or 2 $\mu$m), we mount the substrate onto a mechanical stage with five degree of freedom to adjust position, tilt, and rotation, as shown in Fig.~\ref{Fig:Setup}(a). A cylindrical lens L$_5$ of focal length 25~mm is placed in the path of the ST wave packet, at a distance 30~mm from the lens L$_3$. This lens focuses the synthesized ST wave packet into the film and launches the hybrid guided ST mode. The field at the output is magnified by an objective lens (Olympus UPlanSApo 100x/1.40 oil microscope objective) and imaged to CCD$_{1}$.

\subsection*{Coupling efficiency into the film}

Calculations of the coupling efficiency into the 2-$\mu$m-thick planar waveguide were performed by first calculating its modes at a wavelength of 800~nm taking into consideration its asymmetric index profile. Calibrated measurements of the intensity distribution of the ST wave packets at the input to the film along the waveguide-confined direction $y$ were found to fit well to Gaussian profiles whose widths (full width at half maximum) were in the range $4.8-5.4$~$\mu$m for the six spectral tilt angles used in Fig.~\ref{Fig:ChangingTheta}, and are thus approximately independent of the spectral tilt angle. By calculating the modal overlap, coupling efficiencies into the fundamental TM mode were found to be $38-42\%$, with slightly improved efficiencies occurring at spectral tilt angles near $45^{\circ}$. After correcting for Fresnel losses at the two surfaces, there is an unaccounted loss of $5-10\%$, which we ascribe to all the potential sources of losses in the film (absorption, scattering, coupling to higher-order modes, etc.).

\subsection*{Extracting the group index}

\begin{figure}[t!]
\centering
\includegraphics[width=8.4cm]{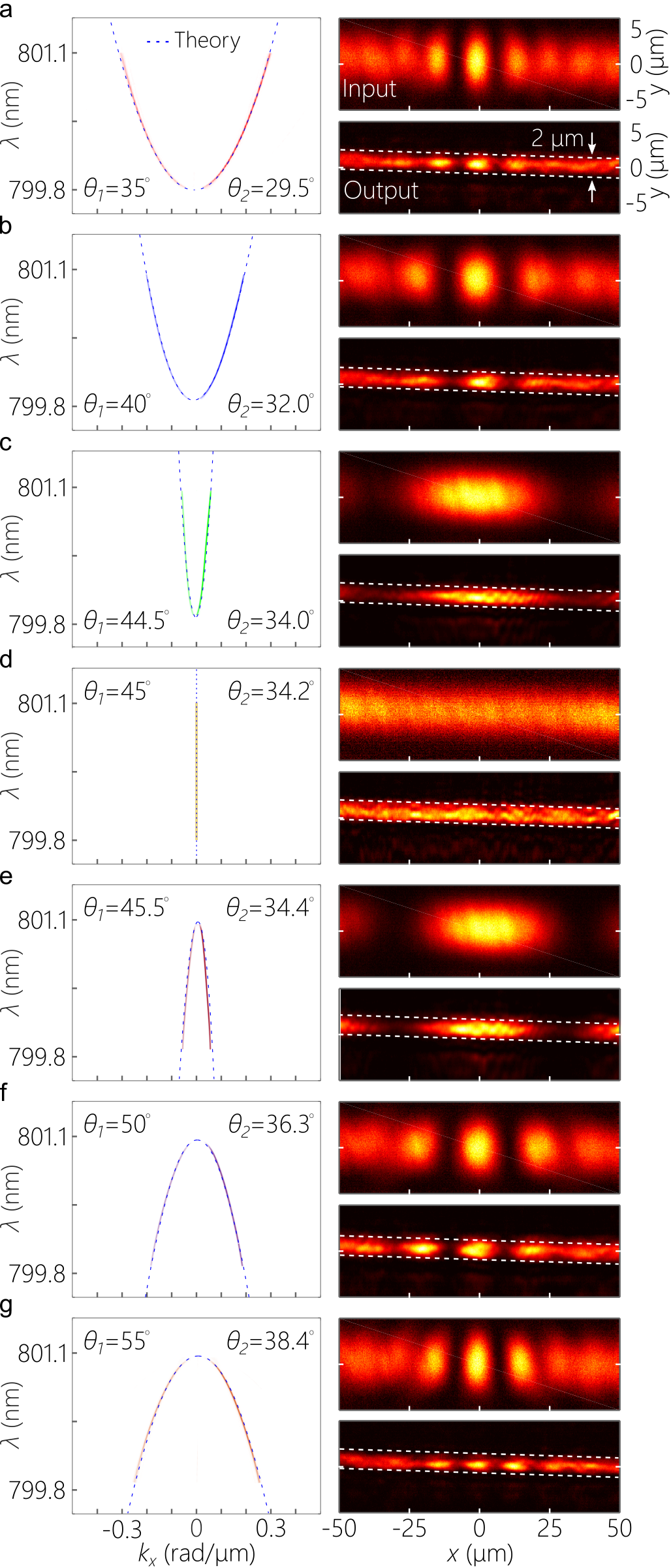}
\caption{Varying the group index of hybrid guided ST modes in a 2-$\mu$m-thick film. (a) The measured spatio-temporal spectral intensity in the $(k_{x},\lambda)$-plane, the intensity $I(x,y)$ at the input to the film and at the output for a spectral tilt angle of $\theta_{1}\!=\!35^{\circ}$, (b) $\theta_{1}\!=\!40^{\circ}$, (c) $\theta_{1}\!=\!44.5^{\circ}$, (d) $\theta_{1}\!=\!45^{\circ}$, (e) $\theta_{1}\!=\!45.5^{\circ}$, (f) $\theta_{1}\!=\!50^{\circ}$, (g) $\theta_{1}\!=\!55^{\circ}$. The dashed curves are theoretical expectations. In each case, a prescribed structure is imposed on the spatio-temporal spectrum by the SLM to realize the targeted value of $\theta_{1}$ and thus $\theta_{2}$ (based on Eq.~\ref{Eq:theta}). The fundamental TM mode is excited along $y$.}
\label{Fig:ChangingTheta}
\end{figure}

\begin{figure}[t!]
\centering
\includegraphics[width=8.8cm]{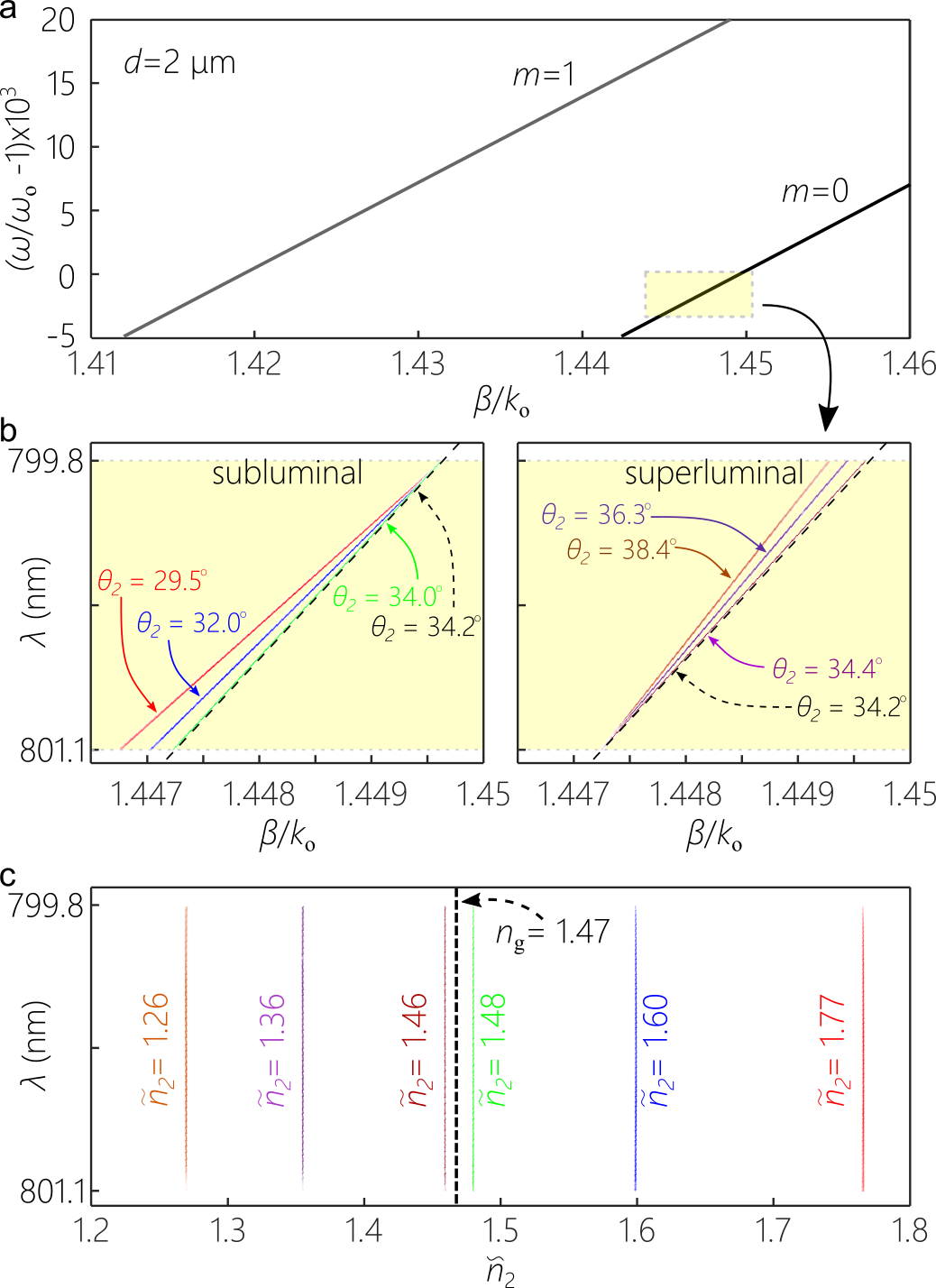}
\caption{ (a) Calculated dispersion relationships for the index-guided planar-waveguide extended modes $m\!=\!0$ and $m\!=\!1$ in a 2-$\mu$m-thick silica film. The highlighted rectangle corresponds to the operational range of our experiment; $\omega_{\mathrm{o}}$ corresponds to a wavelength of 800~nm. (b) Measured dispersion relationships for the hybrid guided ST modes shown in Fig.~\ref{Fig:ChangingTheta}, along with that for the associated planar mode $m\!=\!0$ (dashed). For clarity, the 3 subluminal cases are plotted in the left panel and the 3 superluminal cases are plotted separately in the right panel. The planar waveguide extended mode ($\Delta k_{x}\!=\!0$, corresponding to $\theta_{1}\!=\!45^{\circ}$) is represented by the dashed light-line. (c) Group indices $\widetilde{n}_{2}$ of the hybrid guided ST modes in (b).}
\label{Fig:2um_singlemode}
\end{figure}

Starting from the measured spatio-temporal spectral projection onto the $(k_{x},\lambda)$-plane as plotted in Fig.~\ref{Fig:ChangingTheta}, we obtain the spectral projection onto the $(k_{z},\lambda)$-plane for the hybrid guided ST mode as plotted in Fig.~\ref{Fig:2um_singlemode}(b) \cite{Yessenov19PRA}. The transverse component $k_{y,m}$ in the film is wavelength dependent as is usual for an index-guided mode (rather than the constant value in a waveguide with perfect-mirror boundaries), which we calculate for the fundamental TM planar waveguide mode, and then obtain the axial wave vector component $\beta^{2}\!=\!(n_{2}k_{\mathrm{o}})^{2}-k_{x}^{2}-k_{y,0}^{2}$. The measured values in the $(k_{x},\lambda)$-plane of the incident ST wave packet are mapped to the $(k_{z},\lambda)$-plane of the hybrid guided ST mode through the association of each $k_{x}$ to a corresponding $\beta$.

\clearpage

\clearpage

\clearpage

\clearpage

\clearpage

\clearpage

\clearpage

\end{document}